\documentclass[sigconf]{acmart}
\AtBeginDocument{%
  }

\newcommand{\sysname}{\textsc{BizChat }}
\newcommand{\sysnamenospace}{\textsc{BizChat}}

\copyrightyear{2025}
\acmYear{2025}
\setcopyright{rightsretained}
\acmConference[CHIWORK '25 Adjunct]{CHIWORK '25 Adjunct: Adjunct Proceedings of the 4th Annual Symposium on Human-Computer Interaction for Work}{June 23--25, 2025}{Amsterdam, Netherlands}
\acmBooktitle{CHIWORK '25 Adjunct: Adjunct Proceedings of the 4th Annual Symposium on Human-Computer Interaction for Work (CHIWORK '25 Adjunct), June 23--25, 2025, Amsterdam, Netherlands}
\acmPrice{}
\acmDOI{10.1145/3707640.3731928}
\acmISBN{979-8-4007-1397-2/25/06}

\begin{document}

\title{BizChat: Scaffolding AI-Powered Business Planning for Small Business Owners Across Digital Skill Levels}

\author{Quentin Romero Lauro}
\email{quentinrl@pitt.edu}
\orcid{0009-0000-0620-2146}
\affiliation{%
  \institution{University of Pittsburgh}
  \city{Pittsburgh}
  \state{Pennsylvania}
  \country{USA}
}

\author{Aakash Gautam}
\email{aakash@pitt.edu}
\orcid{0000-0001-8023-4648}
\affiliation{%
  \institution{University of Pittsburgh}
  \city{Pittsburgh}
  \state{Pennsylvania}
  \country{USA}
}

\author{Yasmine Kotturi}
\email{kotturi@umbc.edu}
\orcid{0000-0001-6201-7922}
\affiliation{%
  \institution{Univ of Maryland, Baltimore County}
  \city{Baltimore}
  \state{Maryland}
  \country{USA}
}

\renewcommand{\shortauthors}{Romero Lauro et al.}

\begin{abstract}
Generative AI can help small business owners automate tasks, increase efficiency, and improve their bottom line.
However, despite the seemingly intuitive design of systems like ChatGPT, significant barriers remain for those less comfortable with technology.
To address these disparities, prior work highlights accessory skills---beyond prompt engineering---users must master to successfully adopt generative AI including keyboard shortcuts, editing skills, file conversions, and browser literacy.
Building on a design workshop series and 15 interviews with small businesses, we introduce \sysnamenospace, a large language model (LLM)-powered web application that helps business owners across digital skills levels write their business plan---an essential but often neglected document. 
To do so, \sysnamenospace's interface embodies three design considerations inspired by learning sciences: 
ensuring accessibility to users with less digital skills while maintaining extensibility to power users (``low-floor-high-ceiling''), providing in situ micro-learning to support entrepreneurial education (``just-in-time learning''), and framing interaction around business activities (``contextualized technology introduction''). 
We conclude with plans for a future \sysname deployment.
\end{abstract}


\begin{teaserfigure}
  \centering
  \includegraphics[width=0.95\textwidth]{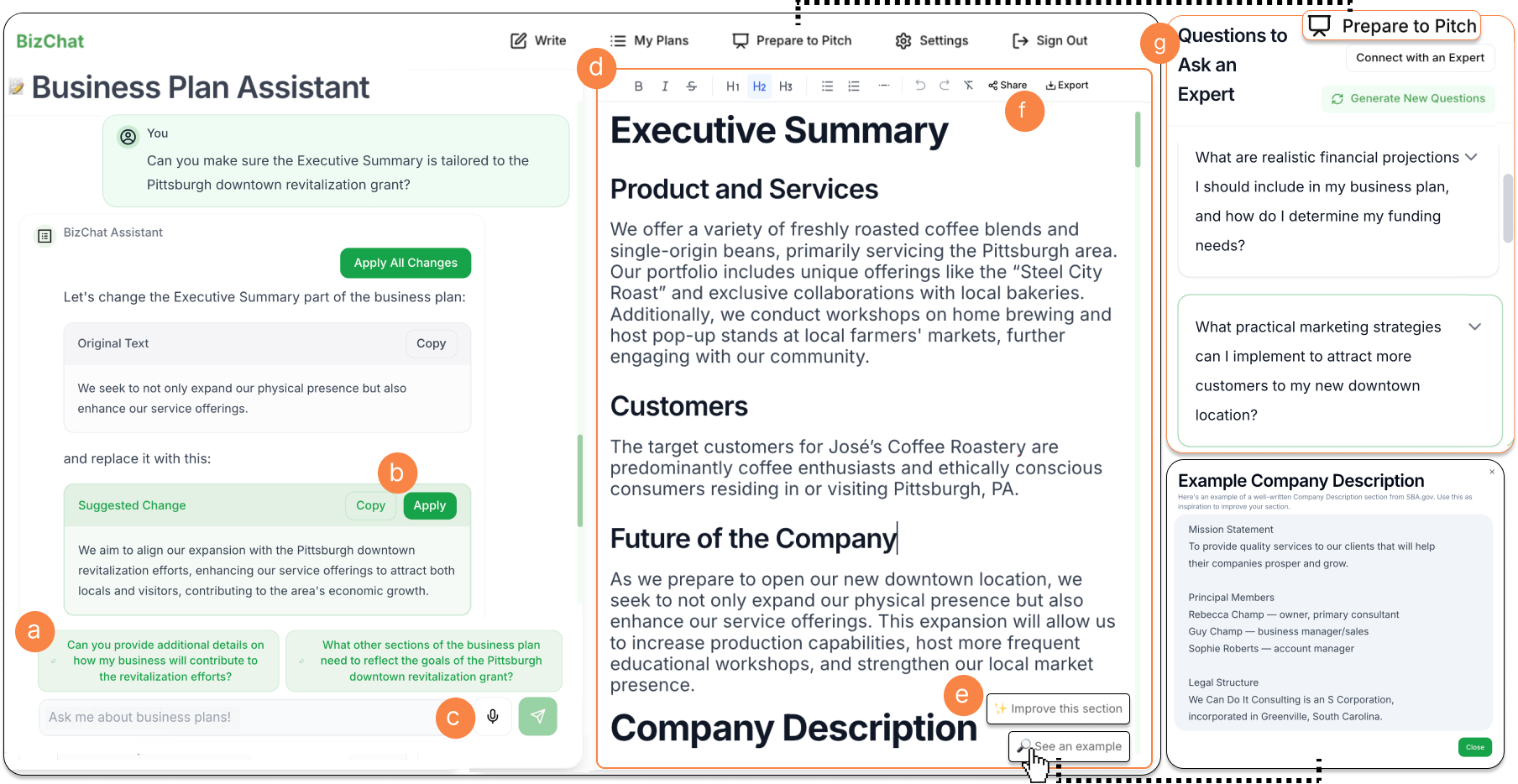}
  \caption{\sysname is a LLM-powered web-app that enables small business owners across digital skill levels to draft business plans from their website or chats, iterate via click-to-apply changes or a rich-text editor, and get feedback from experts---see Section \ref{feature_overview} for details.}
  \Description{BizChat is a system that scaffolds drafting, iterating, and getting feedback on a business plan.}
  \label{fig:teaser}
\end{teaserfigure}

\maketitle

\section{Introduction}
Small business owners are uniquely positioned to reap the benefits of automating tedious back-office tasks with generative AI, given time and resource constraints. 
For instance, large language models (LLMs) can help small business owners to improve their bottom line by automating existing workflows, soliciting and implementing feedback, and unlocking more time for long-term planning~\cite{otis2023uneven}.

Despite the simplicity that generative AI platforms tout~\cite{openai_chatgpt_overview}, integrating generative AI tools into small business workflows is far from straightforward.
For instance, prior work highlights the ``laundry list'' of operational skills beyond prompt engineering, such as browser literacy, password management, cloud storage know-how, and keyboard shortcuts~\cite{kotturi2024deconstructing}, which entrepreneurs must master in order to effectively integrate generative AI into their business.
Additionally, to turn AI-generated outputs into actionable or polished results, entrepreneurs must effectively organize input data, such as converting file types or tailoring content, and refine their outputs to align with their specific business needs~\cite{kotturi2024deconstructing}.
However, current general-purpose AI systems, like ChatGPT or Gemini, presume these tacit knowledge and skill sets among users, creating barriers to adoption and leading to disparities in use between users with and without technical backgrounds~\cite{gusto_2024, pewresearch_2023}.
Further, small business owners who are overly reliant on generative AI---asking for support on tasks that generative AI is not well suited to accomplish---experience exacerbated performance gaps, ultimately magnifying a ``rich gets richer'' dynamic in AI adoption~\cite{otis2023uneven}.

To adopt generative AI, small business owners rely on their social support to create a network of trustworthy information, repurposing single-user ChatGPT accounts to observe how others formulate prompts and thereby reducing costs~\cite{romero2024exploring}.
But this only addresses part of the problem; the systems themselves need improved interfaces and interaction patterns to make this novel technology accessible for users who have diverse levels of technical skills~\cite{weisz2024design}.

In this paper, we extend prior work to investigate how to support entrepreneurs with diverse levels of digital skills to use generative AI systems for their business \cite{kotturi2024deconstructing}. 
In particular, we build on a four-part workshop series
on using generative AI for small businesses with a local entrepreneurial hub dedicated to racial equity in technology in Pittsburgh, Pennsylvania~\cite{kotturi2024deconstructing}.
Through these community-driven workshops and 15 interviews with entrepreneurs, the prior work details the importance of scaffolding actionable use, which pushes beyond hype and provides entrepreneurs with tangible value.
Based on this formative work, we built \sysnamenospace\footnote{\textsc{BizChat} can be freely used at: \url{https://bizchat-io.vercel.app}}: an LLM-powered web application that supports product and service-based entrepreneurs to write business plans---a foundational business document outlining a business's goals and strategies for growth \cite{sba_business_plan}. 
Where Microsoft 365 for Business---and similar commercially available tools---require deep expertise to formulate ``good prompts'' \cite{msf365copilot} to complete amorphous business tasks, \sysname supports entrepreneurs from a range of business and technical backgrounds through scaffolding learning and prompting specifically for business planning. 
Rather than burdening the user with technological fluency, \sysname centers the entrepreneur’s existing knowledge of their business to drive the interaction. 

To inform the interface design, \sysname adheres to three core design considerations drawing from empirical findings from our workshop series and learning science principles: (1) low-floor-high-ceilings~\cite{papert2020mindstorms} to support users with limited digital skills or prior technical knowledge, while offering advanced features for more experienced users; (2) just-in-time learning~\cite{hug2005micro, shail2019using} to facilitate both business and digital skill building for time-constrained entrepreneurs~\cite{avle2019additional}; and (3) contextualizing generative AI within the users’ existing knowledge and goals (i.e., their business), rather than highlighting the novel technology~\cite{bar2007mobile, gautam2020crafting}.
In a future user study and expert evaluation, we aim to explore how \sysnamenospace can facilitate access to generative AI technology for entrepreneurs with diverse digital skills, and examine its subsequent impact on their business operations.

\section{\sysnamenospace's Three Design Considerations}
\label{feature_overview}
In this section, we describe the three design considerations---inspired from learning sciences~\cite{resnick2008falling, novak1999just, hidi2006four,hug2005micro}---which guided \sysnamenospace's interface design and features.

\textbf{(1) Low Floors, High Ceilings.}
\sysnamenospace's first design consideration is to build ``low-floors and high-ceilings''---a pedagogical term which refers to designing interventions that are accessible to beginners while providing opportunities for advanced learners to engage in deeper exploration~\cite{papert2020mindstorms}.
In this context, we argue that a low-floor-high-ceiling design for generative AI systems means creating a system accessible to users with limited digital skills or no prior knowledge of generative AI, while remaining extensible so that entrepreneurs already familiar with tools like ChatGPT will find advanced and useful features.
Prior work revealed that since entrepreneurs have diverse technical and business backgrounds, systems cannot assume knowledge of accessory skills to successfully leverage generative AI tools~\cite{kotturi2024deconstructing}.
As small business owners have diverse levels of typing and writing proficiency---from two-finger typing to touch typing---we scaffold the editing process so that \textbf{Voice-to-Text (\ref{fig:teaser}c)} is available in every step: describing the business, dictating edits, and chatting with \sysnamenospace.
For users comfortable with typing, \sysnamenospace's \textbf{Rich-text Editor (\ref{fig:teaser}d)} enables more control through direct manipulation of their business plan.
As prompting remains a barrier for non-expert users of LLMs~\cite{zamfirescu2023johnny}, \sysname creates two \textbf{Prompt Suggestions (\ref{fig:teaser}a)} for every conversation turn---one prompt focusing on the current topic (exploitation~\cite{pirolli1995information}); the other focusing on a new topic (exploration~\cite{pirolli1995information}). 
When \sysname provides suggestions, the user can one-click \textbf{Apply (\ref{fig:teaser}b)} suggestions from the chat to the business plan in the rich-text editor. 
Simultaneously, for more advanced users, \sysname enables in-line text generation \textbf{(\ref{fig:teaser}e)}, where users can specify criteria to generate text and view exemplars.
Further, the \textbf{Rich-text Editor (\ref{fig:teaser}d)} allows users without word-processing skills to specify text styles and \textbf{Export (\ref{fig:teaser}f)} to a standard template, without the additional barriers of complex document formatting.

\textbf{(2) Just-in-Time Learning.} 
\sysnamenospace's second design consideration is just-in-time learning~\cite{hug2005micro,shail2019using}---in situ micro-learning opportunities that supports entrepreneurial education.
This is especially important because many self-taught entrepreneurs do not have formal business training, nor the time for extended lesson plans~\cite{avle2019additional}---particularly when the value of general-purpose AI is not immediately clear, as participants revealed in our workshop series~\cite{kotturi2024deconstructing}.
With \sysnamenospace, our goal is to help entrepreneurs develop skills in three key areas: understanding what constitutes an effective business plan, building reflection and help-seeking skills, and learning to effectively use generative AI.
To do so, \sysnamenospace's onboarding flow includes an \textbf{Informational Video} reviewing key motivations for business plan creation and detailing their key components.
\sysname then scaffolds users to iterate on each section of their business plan, suggesting relevant \textbf{Examples (\ref{fig:teaser}e)} from U.S. Small Business Administration\footnote{\url{https://www.sba.gov/business-guide/plan-your-business/write-your-business-plan}}.
In addition, \sysname provides users relevant suggested questions as tool-tips in the editor for the \textbf{Business Plan Assistant (\ref{fig:teaser}e)}.
Last, to build help-seeking skills and facilitate self-reflection, \sysnamenospace's \textbf{Prepare to Pitch (\ref{fig:teaser}g)} provides users a list of questions to ask an expert tailored to their business plan and goal which are a helpful preparatory step when approaching business coaches for critical feedback.

\textbf{(3) Contextualized Introduction to Technology.}
General-purpose generative AI tools demand significant sensemaking~\cite{suh2023sensecape, gero2024supporting}, slowing adoption as users struggle to identify practical use cases~\cite{mitnews_2024, voicebot_ai_2023}.
Further, our formative work revealed how entrepreneurs express anxiety about falling behind or ``missing out'' on new technological trends, which compounds hesitation to engage with unfamiliar tools~\cite{kotturi2024deconstructing}. 
To mitigate such anxieties, introducing technology in the context of users’ existing knowledge and goals---rather than focusing on the novel technology---can be a helpful strategy~\cite{bar2007mobile, gautam2020crafting}.
Therefore, \sysnamenospace's third design consideration is to contextualize users' interactions with generative AI within their area of expertise---their business~\cite{hidi2006four}.
For instance, at each conversation turn, the \textbf{Business Plan Assistant (\ref{fig:teaser}b)} suggests changes to the business plan in accordance with the user’s \textbf{Business Plan Goals}, which are explicitly set during onboarding. 
Further, unlike AI systems that present outputs as a completed document, \sysname positions business plans as an evolving document used as a mechanism for planning and growth. 
To this end, \sysname seeks to facilitate deeper engagement with a broader community of support (e.g., expert business coaches, an essential part of an entrepreneurial ecosystem~\cite{hui2020community}).
\sysname offers users to \textbf{Connect with an Expert (\ref{fig:teaser}g)} and aims to decrease reputational risks by providing entrepreneurs a list of \textbf{Questions to Ask an Expert (\ref{fig:teaser}g)} about their business plan.
In doing so, \sysname acknowledges that, even with the latest technology, social support among minority entrepreneurs is essential~\cite{kotturi2024sustaining, kotturi2024peerdea, kotturi2022tech, hui2023community}.

\section{\sysname User Scenario}
To illustrate how \sysname can be used to draft, iterate, and get feedback on a business plan, let us follow José: an aspiring entrepreneur with a coffee roasting venture.
Like many small business owners, José's business success is attributed to his resilience and creativity; he is primarily self-taught and lacks formal training in business or technology.
José wants to expand his business by applying for a local funding opportunity and needs a business plan to be eligible; he finds \sysname linked on the application's resource page. 
\sysname analyzes José's website and generates a draft of his business plan while he watches an informational video about the different sections that constitute an effective business plan. 
Within \sysnamenospace's main interface (Figure \ref{fig:teaser}), he reads his generated draft and notices an error in the Executive Summary section. 
Because he finds his typing slow, he uses voice-to-text to edit the plan within the rich-text editor.
\sysname then recommends improvements to the Market Analysis section.
To verify the changes suggested, José compares an example from SBA.gov\footnotemark[2] with his Market Analysis section.
With José's newfound understanding of what constitutes an effective Market Analysis, he applies the changes.
\sysname then nudges José to seek out expert feedback to continue to iterate on his draft, and provides him with a list of questions to begin the conversation. 
Now, José is ready to submit his business plan, and therefore exports it to a standard template for the city grant application.

\section{\sysname Implementation}
\sysname is a React application built on Next.js, using Firebase to manage data and secure authentication. 
To maintain interactive latencies while achieving well-structured and high-quality business plan generation, each business plan section (e.g., Executive Summary) is generated by asynchronous requests to \texttt{gpt-4-turbo} and then concatenated to create one streamlined business plan. 
We use \texttt{GPT-3.5-turbo} to extract relevant information from web pages and chats during onboarding, which is then used as context to generate the user's first draft business plan.
We use \texttt{GPT-3.5-turbo} for chat, prompt suggestions, and website summaries; \texttt{GPT-4o-mini} for suggestions; \texttt{gpt-4-turbo} for business plans; and \texttt{Whisper-1} for transcription.
Business plan generation prompts are few-shot~\cite{wang2020generalizing} using examples from SBA.gov\footnotemark[2].

\section{Conclusion and Future Work}

Informed by our workshop series, we built \sysname to support small business owners in developing business plans. \sysnamenospace's three design considerations---low floors-high ceilings; just-in-time learning; contextualized introduction of technology---strive to lower barriers to AI adoption for entrepreneurs with diverse digital skills.
Moving forward, to evaluate \sysnamenospace's design, we plan to conduct an in-person workshop series with small business owners to gather qualitative data on usage, followed by a longitudinal deployment study to study key themes in aggregate, and expert evaluation of resulting business plans.
In doing so, we attend to a persistent question within the HCI community amid rapid AI advancements: how can we support users with diverse digital skills to meaningfully engage and shape emerging technologies?



\begin{thebibliography}{30}


\ifx \showCODEN    \undefined \def \showCODEN     #1{\unskip}     \fi
\ifx \showDOI      \undefined \def \showDOI       #1{#1}\fi
\ifx \showISBNx    \undefined \def \showISBNx     #1{\unskip}     \fi
\ifx \showISBNxiii \undefined \def \showISBNxiii  #1{\unskip}     \fi
\ifx \showISSN     \undefined \def \showISSN      #1{\unskip}     \fi
\ifx \showLCCN     \undefined \def \showLCCN      #1{\unskip}     \fi
\ifx \shownote     \undefined \def \shownote      #1{#1}          \fi
\ifx \showarticletitle \undefined \def \showarticletitle #1{#1}   \fi
\ifx \showURL      \undefined \def \showURL       {\relax}        \fi
\providecommand\bibfield[2]{#2}
\providecommand\bibinfo[2]{#2}
\providecommand\natexlab[1]{#1}
\providecommand\showeprint[2][]{arXiv:#2}

\bibitem[Avle et~al\mbox{.}(2019)]%
        {avle2019additional}
\bibfield{author}{\bibinfo{person}{Seyram Avle}, \bibinfo{person}{Julie Hui},
  \bibinfo{person}{Silvia Lindtner}, {and} \bibinfo{person}{Tawanna
  Dillahunt}.} \bibinfo{year}{2019}\natexlab{}.
\newblock \showarticletitle{Additional labors of the entrepreneurial self}.
\newblock \bibinfo{journal}{\emph{Proceedings of the ACM on Human-Computer
  Interaction}} \bibinfo{volume}{3}, \bibinfo{number}{CSCW}
  (\bibinfo{year}{2019}), \bibinfo{pages}{1--24}.
\newblock


\bibitem[Bar et~al\mbox{.}(2007)]%
        {bar2007mobile}
\bibfield{author}{\bibinfo{person}{Fran{\c{c}}ois Bar},
  \bibinfo{person}{Francis Pisani}, {and} \bibinfo{person}{Matthew Weber}.}
  \bibinfo{year}{2007}\natexlab{}.
\newblock \showarticletitle{Mobile technology appropriation in a distant
  mirror: Baroque infiltration, creolization and cannibalism}.
\newblock \bibinfo{journal}{\emph{Seminario sobre Desarrollo Econ{\'o}mico,
  Desarrollo Social y Comunicaciones M{\'o}viles en Am{\'e}rica Latina}}
  (\bibinfo{year}{2007}), \bibinfo{pages}{20--21}.
\newblock


\bibitem[Center(2023)]%
        {pewresearch_2023}
\bibfield{author}{\bibinfo{person}{Pew~Research Center}.}
  \bibinfo{year}{2023}\natexlab{}.
\newblock \bibinfo{booktitle}{\emph{What the Data Says About Americans' Views
  of Artificial Intelligence}}.
\newblock
\urldef\tempurl%
\url{https://www.pewresearch.org/short-reads/2023/11/21/what-the-data-says-about-americans-views-of-artificial-intelligence/}
\showURL{%
\tempurl}
\newblock
\shownote{Accessed: 2025-01-07}.


\bibitem[Gautam et~al\mbox{.}(2020)]%
        {gautam2020crafting}
\bibfield{author}{\bibinfo{person}{Aakash Gautam}, \bibinfo{person}{Deborah
  Tatar}, {and} \bibinfo{person}{Steve Harrison}.}
  \bibinfo{year}{2020}\natexlab{}.
\newblock \showarticletitle{Crafting, communality, and computing: Building on
  existing strengths to support a vulnerable population}. In
  \bibinfo{booktitle}{\emph{Proceedings of the 2020 CHI Conference on Human
  Factors in Computing Systems}}. \bibinfo{pages}{1--14}.
\newblock


\bibitem[Gero et~al\mbox{.}(2024)]%
        {gero2024supporting}
\bibfield{author}{\bibinfo{person}{Katy~Ilonka Gero}, \bibinfo{person}{Chelse
  Swoopes}, \bibinfo{person}{Ziwei Gu}, \bibinfo{person}{Jonathan~K
  Kummerfeld}, {and} \bibinfo{person}{Elena~L Glassman}.}
  \bibinfo{year}{2024}\natexlab{}.
\newblock \showarticletitle{Supporting Sensemaking of Large Language Model
  Outputs at Scale}. In \bibinfo{booktitle}{\emph{Proceedings of the CHI
  Conference on Human Factors in Computing Systems}}. \bibinfo{pages}{1--21}.
\newblock


\bibitem[Gusto(2024)]%
        {gusto_2024}
\bibfield{author}{\bibinfo{person}{Gusto}.} \bibinfo{year}{2024}\natexlab{}.
\newblock \bibinfo{booktitle}{\emph{How SMBs Are Using AI in 2024}}.
\newblock
\urldef\tempurl%
\url{https://gusto.com/company-news/smbs-using-ai-2024#}
\showURL{%
\tempurl}
\newblock
\shownote{Accessed: 2025-01-07}.


\bibitem[Hidi and Renninger(2006)]%
        {hidi2006four}
\bibfield{author}{\bibinfo{person}{Suzanne Hidi} {and} \bibinfo{person}{K~Ann
  Renninger}.} \bibinfo{year}{2006}\natexlab{}.
\newblock \showarticletitle{The four-phase model of interest development}.
\newblock \bibinfo{journal}{\emph{Educational psychologist}}
  \bibinfo{volume}{41}, \bibinfo{number}{2} (\bibinfo{year}{2006}),
  \bibinfo{pages}{111--127}.
\newblock


\bibitem[Hug(2005)]%
        {hug2005micro}
\bibfield{author}{\bibinfo{person}{Theo Hug}.} \bibinfo{year}{2005}\natexlab{}.
\newblock \showarticletitle{Micro Learning and Narration. Exploring
  possibilities of utilization of narrations and storytelling for the designing
  of" micro units" and didactical micro-learning arrangements}. In
  \bibinfo{booktitle}{\emph{fourth Media in Transition conference}},
  Vol.~\bibinfo{volume}{6}. MiT4.
\newblock


\bibitem[Hui et~al\mbox{.}(2020)]%
        {hui2020community}
\bibfield{author}{\bibinfo{person}{Julie Hui}, \bibinfo{person}{Nefer~Ra
  Barber}, \bibinfo{person}{Wendy Casey}, \bibinfo{person}{Suzanne Cleage},
  \bibinfo{person}{Danny~C Dolley}, \bibinfo{person}{Frances Worthy},
  \bibinfo{person}{Kentaro Toyama}, {and} \bibinfo{person}{Tawanna~R
  Dillahunt}.} \bibinfo{year}{2020}\natexlab{}.
\newblock \showarticletitle{Community collectives: Low-tech social support for
  digitally-engaged entrepreneurship}. In \bibinfo{booktitle}{\emph{Proceedings
  of the 2020 CHI conference on human factors in computing systems}}.
  \bibinfo{pages}{1--15}.
\newblock


\bibitem[Hui et~al\mbox{.}(2023)]%
        {hui2023community}
\bibfield{author}{\bibinfo{person}{Julie Hui}, \bibinfo{person}{Kristin
  Seefeldt}, \bibinfo{person}{Christie Baer}, \bibinfo{person}{Lutalo Sanifu},
  \bibinfo{person}{Aaron Jackson}, {and} \bibinfo{person}{Tawanna~R
  Dillahunt}.} \bibinfo{year}{2023}\natexlab{}.
\newblock \showarticletitle{Community Tech Workers: Scaffolding Digital
  Engagement Among Underserved Minority Businesses}.
\newblock \bibinfo{journal}{\emph{Proceedings of the ACM on Human-Computer
  Interaction}} \bibinfo{volume}{7}, \bibinfo{number}{CSCW2}
  (\bibinfo{year}{2023}), \bibinfo{pages}{1--25}.
\newblock


\bibitem[Kotturi et~al\mbox{.}(2024a)]%
        {kotturi2024deconstructing}
\bibfield{author}{\bibinfo{person}{Yasmine Kotturi}, \bibinfo{person}{Angel
  Anderson}, \bibinfo{person}{Glenn Ford}, \bibinfo{person}{Michael Skirpan},
  {and} \bibinfo{person}{Jeffrey~P Bigham}.} \bibinfo{year}{2024}\natexlab{a}.
\newblock \showarticletitle{Deconstructing the Veneer of Simplicity:
  Co-Designing Introductory Generative AI Workshops with Local Entrepreneurs}.
  In \bibinfo{booktitle}{\emph{Proceedings of the CHI Conference on Human
  Factors in Computing Systems}}. \bibinfo{pages}{1--16}.
\newblock


\bibitem[Kotturi et~al\mbox{.}(2024b)]%
        {kotturi2024sustaining}
\bibfield{author}{\bibinfo{person}{Yasmine Kotturi}, \bibinfo{person}{Julie
  Hui}, \bibinfo{person}{TJ Johnson}, \bibinfo{person}{Lutalo Sanifu}, {and}
  \bibinfo{person}{Tawanna~R Dillahunt}.} \bibinfo{year}{2024}\natexlab{b}.
\newblock \showarticletitle{Sustaining Community-Based Research in Computing:
  Lessons from Two Tech Capacity Building Initiatives for Local Businesses}.
\newblock \bibinfo{journal}{\emph{Proceedings of the ACM on Human-Computer
  Interaction}} \bibinfo{volume}{8}, \bibinfo{number}{CSCW1}
  (\bibinfo{year}{2024}), \bibinfo{pages}{1--31}.
\newblock


\bibitem[Kotturi et~al\mbox{.}(2022)]%
        {kotturi2022tech}
\bibfield{author}{\bibinfo{person}{Yasmine Kotturi}, \bibinfo{person}{Herman~T
  Johnson}, \bibinfo{person}{Michael Skirpan}, \bibinfo{person}{Sarah~E Fox},
  \bibinfo{person}{Jeffrey~P Bigham}, {and} \bibinfo{person}{Amy Pavel}.}
  \bibinfo{year}{2022}\natexlab{}.
\newblock \showarticletitle{Tech help desk: Support for local entrepreneurs
  addressing the Long Tail of computing challenges}. In
  \bibinfo{booktitle}{\emph{Proceedings of the 2022 CHI Conference on Human
  Factors in Computing Systems}}. \bibinfo{pages}{1--15}.
\newblock


\bibitem[Kotturi et~al\mbox{.}(2024c)]%
        {kotturi2024peerdea}
\bibfield{author}{\bibinfo{person}{Yasmine Kotturi}, \bibinfo{person}{Jenny
  Yu}, \bibinfo{person}{Pranav Khadpe}, \bibinfo{person}{Erin Gatz},
  \bibinfo{person}{Harvey Zheng}, \bibinfo{person}{Sarah~E Fox}, {and}
  \bibinfo{person}{Chinmay Kulkarni}.} \bibinfo{year}{2024}\natexlab{c}.
\newblock \showarticletitle{Peerdea: Co-Designing a Peer Support Platform with
  Creative Entrepreneurs}.
\newblock \bibinfo{journal}{\emph{Proceedings of the ACM on Human-Computer
  Interaction}} \bibinfo{volume}{8}, \bibinfo{number}{CSCW1}
  (\bibinfo{year}{2024}), \bibinfo{pages}{1--24}.
\newblock


\bibitem[{Microsoft}(2025)]%
        {msf365copilot}
\bibfield{author}{\bibinfo{person}{{Microsoft}}.}
  \bibinfo{year}{2025}\natexlab{}.
\newblock \bibinfo{title}{Prompts to try}.
\newblock
\newblock
\urldef\tempurl%
\url{https://support.microsoft.com/en-us/copilot-microsoft365-chat}
\showURL{%
\tempurl}
\newblock
\shownote{Accessed: 2025-04-27}.


\bibitem[News(2024)]%
        {mitnews_2024}
\bibfield{author}{\bibinfo{person}{MIT News}.} \bibinfo{year}{2024}\natexlab{}.
\newblock \bibinfo{booktitle}{\emph{Reasoning Skills of Large Language Models
  Are Often Overestimated}}.
\newblock
\urldef\tempurl%
\url{https://news.mit.edu/2024/reasoning-skills-large-language-models-often-overestimated-0711?utm_source=chatgpt.com}
\showURL{%
\tempurl}
\newblock
\shownote{Accessed: 2025-01-07}.


\bibitem[Novak et~al\mbox{.}(1999)]%
        {novak1999just}
\bibfield{author}{\bibinfo{person}{Gregor~M Novak}, \bibinfo{person}{Evelyn~T
  Patterson}, \bibinfo{person}{Andrew~D Gavrin}, {and}
  \bibinfo{person}{Wolfgang Christian}.} \bibinfo{year}{1999}\natexlab{}.
\newblock \showarticletitle{Just-in-time teaching blending active learning with
  web technology}.
\newblock  (\bibinfo{year}{1999}).
\newblock


\bibitem[{OpenAI}(2025)]%
        {openai_chatgpt_overview}
\bibfield{author}{\bibinfo{person}{{OpenAI}}.} \bibinfo{year}{2025}\natexlab{}.
\newblock \bibinfo{title}{ChatGPT Overview}.
\newblock
\newblock
\urldef\tempurl%
\url{https://openai.com/chatgpt/overview/}
\showURL{%
\tempurl}
\newblock
\shownote{Accessed: 2025-01-03}.


\bibitem[Otis et~al\mbox{.}(2023)]%
        {otis2023uneven}
\bibfield{author}{\bibinfo{person}{Nicholas Otis}, \bibinfo{person}{Rowan~P
  Clarke}, \bibinfo{person}{Solene Delecourt}, \bibinfo{person}{David Holtz},
  {and} \bibinfo{person}{Rembrand Koning}.} \bibinfo{year}{2023}\natexlab{}.
\newblock \showarticletitle{The uneven impact of generative AI on
  entrepreneurial performance}.
\newblock \bibinfo{journal}{\emph{Available at SSRN 4671369}}
  (\bibinfo{year}{2023}).
\newblock


\bibitem[Papert(2020)]%
        {papert2020mindstorms}
\bibfield{author}{\bibinfo{person}{Seymour~A Papert}.}
  \bibinfo{year}{2020}\natexlab{}.
\newblock \bibinfo{booktitle}{\emph{Mindstorms: Children, computers, and
  powerful ideas}}.
\newblock \bibinfo{publisher}{Basic books}.
\newblock


\bibitem[Pirolli and Card(1995)]%
        {pirolli1995information}
\bibfield{author}{\bibinfo{person}{Peter Pirolli} {and} \bibinfo{person}{Stuart
  Card}.} \bibinfo{year}{1995}\natexlab{}.
\newblock \showarticletitle{Information foraging in information access
  environments}. In \bibinfo{booktitle}{\emph{Proceedings of the SIGCHI
  conference on Human factors in computing systems}}. \bibinfo{pages}{51--58}.
\newblock


\bibitem[Resnick(2008)]%
        {resnick2008falling}
\bibfield{author}{\bibinfo{person}{Mitchel Resnick}.}
  \bibinfo{year}{2008}\natexlab{}.
\newblock \showarticletitle{Falling in love with Seymour’s ideas}. In
  \bibinfo{booktitle}{\emph{American Educational Research Association (AERA)
  annual conference}}.
\newblock


\bibitem[Romero~Lauro et~al\mbox{.}(2024)]%
        {romero2024exploring}
\bibfield{author}{\bibinfo{person}{Quentin Romero~Lauro},
  \bibinfo{person}{Jeffrey~P Bigham}, {and} \bibinfo{person}{Yasmine Kotturi}.}
  \bibinfo{year}{2024}\natexlab{}.
\newblock \showarticletitle{Exploring the Role of Social Support When
  Integrating Generative AI in Small Business Workflows}. In
  \bibinfo{booktitle}{\emph{Companion Publication of the 2024 Conference on
  Computer-Supported Cooperative Work and Social Computing}}.
  \bibinfo{pages}{485--492}.
\newblock


\bibitem[Shail(2019)]%
        {shail2019using}
\bibfield{author}{\bibinfo{person}{Mrigank~S Shail}.}
  \bibinfo{year}{2019}\natexlab{}.
\newblock \showarticletitle{Using micro-learning on mobile applications to
  increase knowledge retention and work performance: a review of literature}.
\newblock \bibinfo{journal}{\emph{Cureus}} \bibinfo{volume}{11},
  \bibinfo{number}{8} (\bibinfo{year}{2019}).
\newblock


\bibitem[Suh et~al\mbox{.}(2023)]%
        {suh2023sensecape}
\bibfield{author}{\bibinfo{person}{Sangho Suh}, \bibinfo{person}{Bryan Min},
  \bibinfo{person}{Srishti Palani}, {and} \bibinfo{person}{Haijun Xia}.}
  \bibinfo{year}{2023}\natexlab{}.
\newblock \showarticletitle{Sensecape: Enabling multilevel exploration and
  sensemaking with large language models}. In
  \bibinfo{booktitle}{\emph{Proceedings of the 36th Annual ACM Symposium on
  User Interface Software and Technology}}. \bibinfo{pages}{1--18}.
\newblock


\bibitem[{U.S. Small Business Administration}(nd)]%
        {sba_business_plan}
\bibfield{author}{\bibinfo{person}{{U.S. Small Business Administration}}.}
  \bibinfo{year}{n.d.}\natexlab{}.
\newblock \bibinfo{title}{Write Your Business Plan}.
\newblock
\newblock
\urldef\tempurl%
\url{https://www.sba.gov/business-guide/plan-your-business/write-your-business-plan}
\showURL{%
\tempurl}
\newblock
\shownote{Accessed: 2025-01-07}.


\bibitem[Voicebot.ai(2023)]%
        {voicebot_ai_2023}
\bibfield{author}{\bibinfo{person}{Voicebot.ai}.}
  \bibinfo{year}{2023}\natexlab{}.
\newblock \bibinfo{booktitle}{\emph{Generative {AI} Apps Struggle with
  Retention and Engagement [Charts]}}.
\newblock
\urldef\tempurl%
\url{https://voicebot.ai/2023/10/03/generative-ai-apps-struggle-with-retention-and-engagement-charts/}
\showURL{%
\tempurl}
\newblock
\shownote{Accessed: 2025-01-07}.


\bibitem[Wang et~al\mbox{.}(2020)]%
        {wang2020generalizing}
\bibfield{author}{\bibinfo{person}{Yaqing Wang}, \bibinfo{person}{Quanming
  Yao}, \bibinfo{person}{James~T Kwok}, {and} \bibinfo{person}{Lionel~M Ni}.}
  \bibinfo{year}{2020}\natexlab{}.
\newblock \showarticletitle{Generalizing from a few examples: A survey on
  few-shot learning}.
\newblock \bibinfo{journal}{\emph{ACM computing surveys (csur)}}
  \bibinfo{volume}{53}, \bibinfo{number}{3} (\bibinfo{year}{2020}),
  \bibinfo{pages}{1--34}.
\newblock


\bibitem[Weisz et~al\mbox{.}(2024)]%
        {weisz2024design}
\bibfield{author}{\bibinfo{person}{Justin~D Weisz}, \bibinfo{person}{Jessica
  He}, \bibinfo{person}{Michael Muller}, \bibinfo{person}{Gabriela Hoefer},
  \bibinfo{person}{Rachel Miles}, {and} \bibinfo{person}{Werner Geyer}.}
  \bibinfo{year}{2024}\natexlab{}.
\newblock \showarticletitle{Design Principles for Generative AI Applications}.
  In \bibinfo{booktitle}{\emph{Proceedings of the CHI Conference on Human
  Factors in Computing Systems}}. \bibinfo{pages}{1--22}.
\newblock


\bibitem[Zamfirescu-Pereira et~al\mbox{.}(2023)]%
        {zamfirescu2023johnny}
\bibfield{author}{\bibinfo{person}{JD Zamfirescu-Pereira},
  \bibinfo{person}{Richmond~Y Wong}, \bibinfo{person}{Bjoern Hartmann}, {and}
  \bibinfo{person}{Qian Yang}.} \bibinfo{year}{2023}\natexlab{}.
\newblock \showarticletitle{Why Johnny can’t prompt: how non-AI experts try
  (and fail) to design LLM prompts}. In \bibinfo{booktitle}{\emph{Proceedings
  of the 2023 CHI Conference on Human Factors in Computing Systems}}.
  \bibinfo{pages}{1--21}.
\newblock


\end{thebibliography}
\end{document}